\documentclass[prx,aps,superscriptaddress,twocolumn]{revtex4-2}

\usepackage{dsfont}
\usepackage{mathtext}
\usepackage[pdftex]{graphicx}
\usepackage{mathtools}
\usepackage[usenames,dvipsnames]{xcolor}
\usepackage{amsmath}
\usepackage{amssymb,mathrsfs}
\usepackage{graphics}
\usepackage{epstopdf}
\usepackage{mathtext}
\usepackage{yfonts}
\usepackage{bm}
\usepackage{epstopdf}

\usepackage{color}
\usepackage[colorlinks=true,linkcolor=blue]{hyperref}

\begin{document}

\title{Entanglement generation in capacitively coupled Transmon-cavity system}


\author{Jian-Zhuang Wu}
\affiliation{School of Science, Inner Mongolia University of Science and Technology, Baotou 014010, People's Republic of China}

\author{Lian-E Lu}
\affiliation{School of Science, Inner Mongolia University of Science and Technology, Baotou 014010, People's Republic of China}

\author{Xinyu Zhao}
\email[Correspondence email address: ]{xzhao@fzu.edu.cn}%
\affiliation{Department of Physics, Fuzhou University, Fuzhou 350116, People's Republic of China}

\author{Yong-Hong Ma}
\email[Correspondence email address: ]{myhdlut@126.com}
\affiliation{School of Science, Inner Mongolia University of Science and Technology, Baotou 014010, People's Republic of China}

\begin{abstract}

In this paper, the higher energy levels of the transmon qubit are taken into consideration to investigate the continuous variable entanglement generation between the transmon qubit and the single-mode cavity. Based on the framework of cavity quantum electrodynamics, we show the entanglement generation depends on the the driving field intensity, coupling strength, cavity field frequency, and qubit frequency. The numerical results show that strong entanglement can be generated by properly tuning these parameters. It is our hope that the results presented in this paper may lead to a better understanding of quantum entanglement generation in cavity QED system and provide new perspectives for further research in quantum information processing.


\end{abstract}

\maketitle

\section{Introduction}\label{sec1}
Quantum entanglement is one of the fundamental properties of quantum
mechanics ~\cite{win4,hor9,mia9,man2,hua1,hua9,roc0}, and it has been demonstrated to be a key resource for quantum information processing tasks. As a unique feature of quantum mechanics, quantum entanglement is often observed between microscopic systems such as photons, ions, and atoms. Manipulating and generating entanglement in mesoscopic or even macroscopic systems is an extremely valuable research topic. On one hand, it proves that quantum mechanics in general is applicable to arbitrary systems, either microscopic or macroscopic. On the other hand, it may reveal the boundary between quantum and classical realms, and further reveal the transition from classical to quantum realms.

In various fields related to quantum entanglement ~\cite{may2,bro1,lia1,soh0,zha9,zha3,pin5,huan9,zhe0}, the interaction model between cavity field and qubits has received widespread attention because it has potential in both practical applications and theoretical studies \cite{maprb,gor0,gio1}. Specifically, it shows huge potential to be the hopeful candidate of the building blocks of a quantum computer, the qubit. The interaction model between cavity field and Transmon has been demonstrated for large-scale quantum computing and even more complex structures such as quantum networks and quantum signal processing \cite{gen8}. There are several advantages for the interaction model between Transmon and cavity. In recent experiments, quantum supremacy (quantum computational advantage) is achieved in this system, where the computational efficiency may surpass any classical computers in solving some particular problems. Besides, the Transmon and cavity system is scalable: the interaction model between cavity field and Transmon is easy to scale for performance improvements by simply adding more qubits to tackle more complex computational problems \cite{ved4,gen9,ian8,ham9,wal0,zho1}. Furthermore, it has a wide range of applications: the interaction model between cavity field and Transmon has been widely used in quantum computing, quantum communication, quantum simulation, and even in fields beyond \cite{gor0,gio1,bla1,leg5}. Therefore, understanding and controlling entanglement in such a system is crucial for quantum information science.

In the field of quantum information processing, two-level quantum systems or qubits are commonly used. However, expanding the Hilbert space dimensions to a d-level system, also known as a "multilevel quantum bit," can offer significant computational advantages. Research on hybrid systems combining superconducting circuits with other quantum systems and solid-state devices is ongoing \cite{tao5}. For instance, Peterer et al. \cite{pet5} explored the energy decay and phase coherence of the first five energy levels of a Transmon qubit embedded in a threedimensional cavity, suggesting the feasibility of encoding quantum information in higher energy levels of Transmon qubits. Goss et al. \cite{goss2} characterized the differential AC Stark shift on two fixed-frequency three-level transmon qubits, using it to generate dynamic three-level entangled phases. However, the continuous-variable entanglement of the Transmon and cavity coupled system has yet been investigated.

In our study, we consider the continuous variable entanglement generation between the transmon qubit and the single-mode cavity.There are several factors that determine the entanglement generation between Transmon and cavity fields, including driving field strength, coupling constants, cavity field frequency, and Transmon frequency. These parameters play crucial roles in controlling the entanglement between Transmon and cavity fields, as well as improving the efficiency and accuracy of quantum information processing. Based on the full Hamiltonian and the master equation, we analyze the entanglement generation between cavity field and Transmon, and study the impact of these parameters on quantum entanglement generation and duration through numerical simulations. Our goal is to gain a deeper understanding of how these parameters collectively affect the entanglement between Transmon and the cavity field, providing a theoretical foundation for more efficient manipulation of quantum entanglement in experiments.


\section{Physical Model and Dynamical Equations}

We consider a Transmon-cavity model, where the Hamiltonian for the capacitively coupled Transmon-cavity \cite{bla1} can be described as (assuming $\hbar=1$).
\begin{equation}
\begin{split}
\hat{H}=&\hbar{{\omega}_{r}}{{\hat{a}}^{\dagger}}\hat{a}+\hbar{{\omega}_{q}}{{\hat{b}}^{\dagger}}\hat{b}-\frac{{{E}_{c}}}{2}{{\hat{b}}^{\dagger}}{{\hat{b}}^{\dagger}}\hat{b}\hat{b}\\
&+\hbar g\left({{\hat{b}}^{\dagger}}\hat{a}+\hat{b}{{\hat{a}}^{\dagger}}\right)+i\hbar E\left({{e}^{-i{{\omega}_{0}}t}}{{\hat{a}}^{\dagger}}-{{e}^{i{{\omega}_{0}}t}}\hat{a}\right).
\end{split}
\label{eq:H}
\end{equation}

\noindent Equation~(\ref{eq:H}) describes the exchange of microwave photons between Transmon and cavity, where the exchange is virtual photon exchange. This system is commonly used in superconducting quantum computing to harness the properties of quantum entanglement. In Eq.~(\ref{eq:H}), the first item is the Hamiltonian of the resonant cavity. ${{\hat{a}}^{\dagger}}$ and $\hat{a}$ are operators for creating and annihilating the cavity field, respectively, while ${{\omega}_{r}}$ represents the frequency of the cavity field. The second term is the Hamiltonian of Transmon qubits. ${{\hat{b}}^{\dagger}}$ and $\hat{b}$ are the creation and annihilation operators of qubits, and ${{\omega}_{q}}$ is the frequency of qubits. The third term is the anharmonicity of Transmon. Due to the energy levels of superconducting qubits not being equally spaced, this term describes such nonlinearity, where ${{E}_{c}}$ represents charge energy. The fourth term is the coupling between Transmon and cavity fields, where $g$ represents the coupling strength and parameter $g=\omega_{r} \frac{C_{g}}{C_{\sum}}(\frac{E_{J}}{2E_{C}})^{\frac{1}{4}}\sqrt{\frac{\pi Z_{r}}{R_{k}}}$\cite{bla1}. This term describes the interaction between Transmon and cavity fields. The last term describes a driven by an external force, where $E$ is the driving strength and parameter $E=\sqrt{\frac{2P}{\hbar{{\omega}_{0}}}}$ is related to input laser power $P$. This item describes how external driving changes the state of the cavity field.

The form of this Hamiltonian is very common in quantum optics and superconducting quantum computing. By selecting appropriate parameters, various interesting physical effects can be achieved, such as quantum entanglement, quantum logic gates, and quantum error correction.

Rotating into the interaction picture, the Hamiltonian can be rewritten as:
\begin{equation}
\begin{split}
\hat{H}=&\hbar{{\omega}_{a}}{{\hat{a}}^{\dagger}}\hat{a}+\hbar{{\omega}_{b}}{{\hat{b}}^{\dagger}}\hat{b}-\frac{{{E}_{c}}}{2}{{\hat{b}}^{\dagger}}{{\hat{b}}^{\dagger}}\hat{b}\hat{b}+\hbar g\left({{\hat{b}}^{\dagger}}\hat{a}+\hat{b}{{\hat{a}}^{\dagger}}\right)\\
&+i\hbar E\left({{\hat{a}}^{\dagger}}-\hat{a}\right),
\end{split}
\end{equation}

\noindent where $\omega_{a}=\omega_{r}-\omega_{0}$,$\omega_{b}=\omega_{q}-\omega_{0}$.
Using the Heisenberg equation of motion, one can obtain the time evolution of the cavity field and vibrating end mirror from the above Hamiltonian. In the absence of damping and noise, the dynamics of cavity modes and qubits can be described by a set of coupled first-order differential equations known as quantum Langevin equations, which are given in Appendix A. Combining all the Langevin equations, the dynamical equations can be written in a very compact form as
\begin{eqnarray}
\dot{f}(t)=Af(t)+\chi(t),\label{eq:f}
\end{eqnarray}

\noindent where $f(t)$ is a vector containing sixteen mean values of different operators.
\begin{equation*}
\begin{split}
f(t)^{T}=(&\langle \hat{a}\rangle,\langle{\hat{a}^{\dagger}}\rangle,\langle \hat{b}\rangle,\langle{\hat{b}^{\dagger}}\rangle,\langle{\hat{a}\hat{a}}\rangle,\langle{\hat{a}^{\dagger}}\hat{a}^{\dagger}\rangle,\langle{\hat{b}\hat{b}}\rangle,\langle{\hat{b}^{\dagger}}\hat{b}^{\dagger}\rangle,\langle \hat{a}\hat{b}\rangle,\\
&\langle{\hat{a}^{\dagger}}{\hat{b}^{\dagger}}\rangle,\langle{\hat{a}^{\dagger}}\hat{b}\rangle,\langle \hat{a}{\hat{b}^{\dagger}}\rangle,\langle{\hat{a}^{\dagger}}\hat{a}\rangle,\langle \hat{a}{\hat{a}^{\dagger}}\rangle,\langle{\hat{b}^{\dagger}}\hat{b}\rangle,\langle \hat{b}{\hat{b}^{\dagger}}\rangle)
\end{split}
\end{equation*}

\noindent The vector $\chi(t)^{T}=(E,E,0,0,0,0,0,0,0,0,0,0,0,0,0,0)$ is the constant term in the matrix equation, and the coefficient matrix $A$ in Eq.~(\ref{eq:f}) can be found in Appendix A.


\section{Results and Discussion}


Our goal is to study the possible quantum entanglement between Transmon and cavity under given conditions. The quantum entanglement in such system is quantified by the logarithmic negativity ${{E}_{N}}$ \cite{ple5}, which is defined as
\begin{eqnarray}
{{E}_{N}}=\max\left[0,-{{\ln}}{2{v}_{-}}\right],
\end{eqnarray}

\noindent where ${{v}_{-}}=\sqrt{\left(\Delta\left(\Gamma\right)-\sqrt{\Delta{{\left(\Gamma\right)}^{2}}-4Det\Gamma}\right)/2}$ is the smallest eigenvalue of the covariance matrix $\Gamma$. The function $\Delta\left(\Gamma\right)$ is defined as $\Delta\left(\Gamma\right)=\det A+\det B-2\det C$, and the corresponding covariance matrix $\Gamma$ is:
\begin{align}
\Gamma=\begin{bmatrix}A & C\\
C^{T} & B
\end{bmatrix}
\end{align}

\noindent In the equation above, the covariance matrix $\Gamma$ is a $4\times4$ matrix that we have partitioned into a $2\times2$ matrix. Each element of this matrix is denoted by ${{\Gamma}_{ij}}$.
\begin{align}
\Gamma_{ij}=\frac{1}{2}\left\langle x_{i}x_{j}+x_{j}x_{i}\right\rangle -\left\langle x_{i}\right\rangle \left\langle x_{j}\right\rangle \label{eq:Gamma}
\end{align}

\noindent The operators in Eq.~(\ref{eq:Gamma}) can be replaced as linear combinations of operators $\hat{a}$, $\hat{b}$, $\hat{a}^{\dagger}$, and $\hat{b}^{\dagger}$ as $x_{1}=\frac{\hat{a}+{{\hat{a}}^{\dagger}}}{\sqrt{2}}$, $x_{2}=\frac{\hat{a}-{{\hat{a}}^{\dagger}}}{\sqrt{2}i}$, $x_{3}=\frac{\hat{b}+{{\hat{b}}^{\dagger}}}{\sqrt{2}}$, and $x_{4}=\frac{\hat{b}-{{\hat{b}}^{\dagger}}}{\sqrt{2}i}$. Therefore, the covariance matrix $\Gamma$ can be computed by using the equations in Appendix B.

\begin{figure*}[tp]
\centering
\includegraphics[width=1\linewidth]{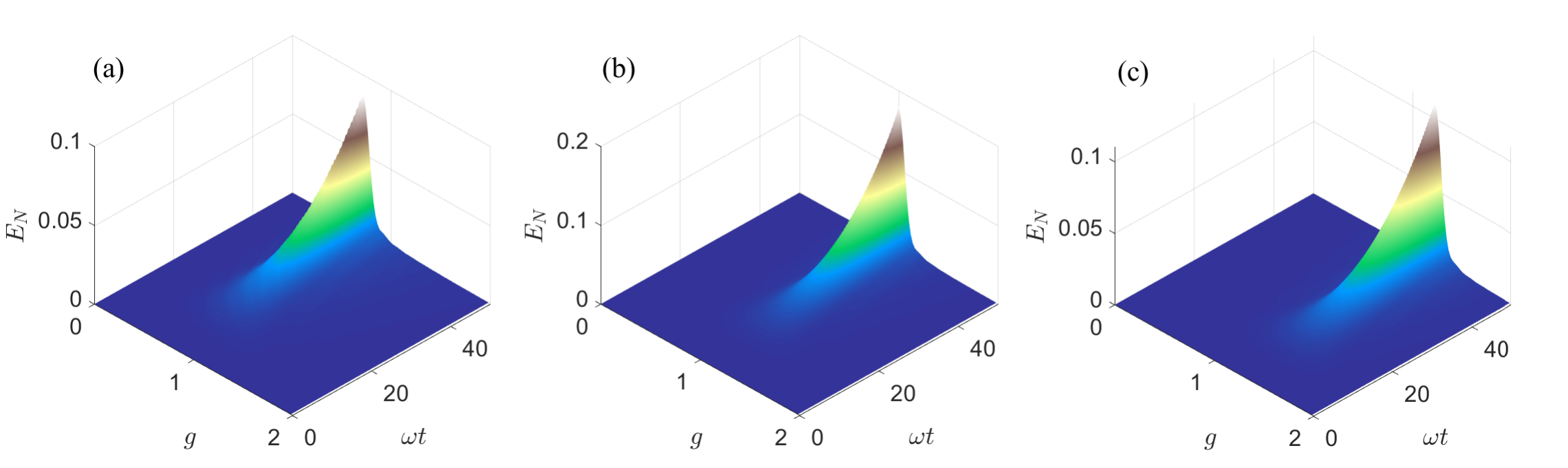}

\caption{Time evolution of entanglement measured by $E_{N}$. For different coupling constants $g$, the generation of $E_{N}$ is quite different. In each subplot, we choose different $\omega_{a}$ as (a) $\omega_{a}=0.5$, (b) $\omega_{a}=1$, (c) $\omega_{a}=1.5$. The other parameters are $\omega_{b}=\omega=1$, $E=0.1$,$\gamma_s=5\times10^{-3}$.}
\label{fig1}
\end{figure*}

To observe the entanglement generation, we set the initial state to be the vacuum state. With the help of numerical calculations, we have obtained the time evolution of Negativity $E_{N}$ for Transmon qubits and cavity systems under different initial conditions.In the numerical calculations, the following system parameters are taken:$\omega_{0}=5.46MHz$, $g/{\omega_{r}}=0.1~0.5$, $\omega_{r}/{2\pi}=3~11GHz$, $\omega_{q}/{2{\pi}}=3~11GHz$, $\gamma_{s}=0.0001$,which are based on current experiment conditions \cite{koc7,pan5,koc3,hu7,mar3,ase4,bar4,rea6}.

\begin{figure}[t]
\centering
\includegraphics[width=1.00\linewidth]{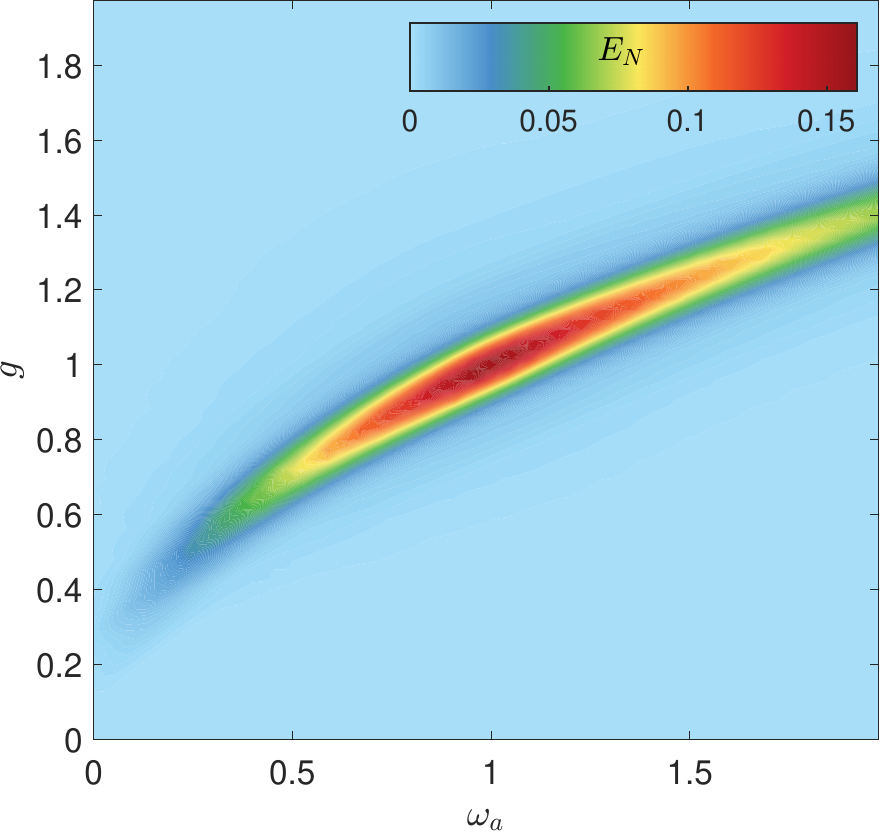}
\caption{Entanglement at $\omega t=50$ for different coupling constants $g$ and $\omega_{a}$. the qubit frequency is fixed as $\omega_{b}=1$. The other parameters are $E=0.1$ and $\gamma_s=5\times10^{-3}$.}
\label{fig2}
\end{figure}

Figure \ref{fig1} shows the time evolution of $E_{N}$ for different coupling constants g when the detuning between cavity field and driving field $\omega_{a}$ is set to 0.5, 1, and 1.5 respectively. From all three subplots in Fig. \ref{fig1}, the entanglement dynamics are similar. There will be always a critical value $g_{c}$, when $g\approx g_{c}$, the entanglement generation is increasing exponentially fast. Except this value, when $g$ is far from $g_{c}$, the entanglement is increasing much slower. The critical value $g_{c}$ obviously depends on the choice of $\omega_{a}$. To be specific, from Fig. \ref{fig1}(a), it can be observed that when $\omega_{a}$ is set to 0.5, there exists entanglement between the system and $E_{N}$ reaches its maximum value between 0.05 and 0.1 when $g$ is around 0.7. From Figure \ref{fig1}(b), it can be seen that when $\omega_{a}$ is set to 1, entanglement $E_{N}$ reaching its maximum value between 0.1 and 0.2 when $g$ is around 1. From Fig. \ref{fig1}(c), when $\omega_{a}$ is set to 1.5, the strongest entanglement generation reach its maximum value 0.11 at $g$ is around 1.3. It is worth to note that $E_{N}$ reaches its highest value during time evolution when the two detunings $\omega_{a}$ and $\omega_{b}$ are equal. We believe the resonance enhance the entanglement in the system in this case.

As we have pointed out in Fig. \ref{fig2}, one can observe that as $\omega_{a}$ increases, the peak of Negativity $E_{N}$ curve moves towards larger coupling constant $g$. In order to better characterize this feature, we plot the maximum value of $E_{N}$ in Fig. \ref{fig2} to get a closer look at the relation between $\omega_{a}$ and $g$. From Fig. \ref{fig2}, max $E_{N}$ appears in a curve in the parameter space spanned in $g$ and $\omega_{a}$. The shape of the curve is clearly shown in Fig. \ref{fig2}. In the region $\omega_{a}$ is large, it is roughly a linear curve, while in the region $\omega_{a}\rightarrow0$, it drops rather fast.

\begin{figure}[t]
\centering
\includegraphics[width=1.00\linewidth]{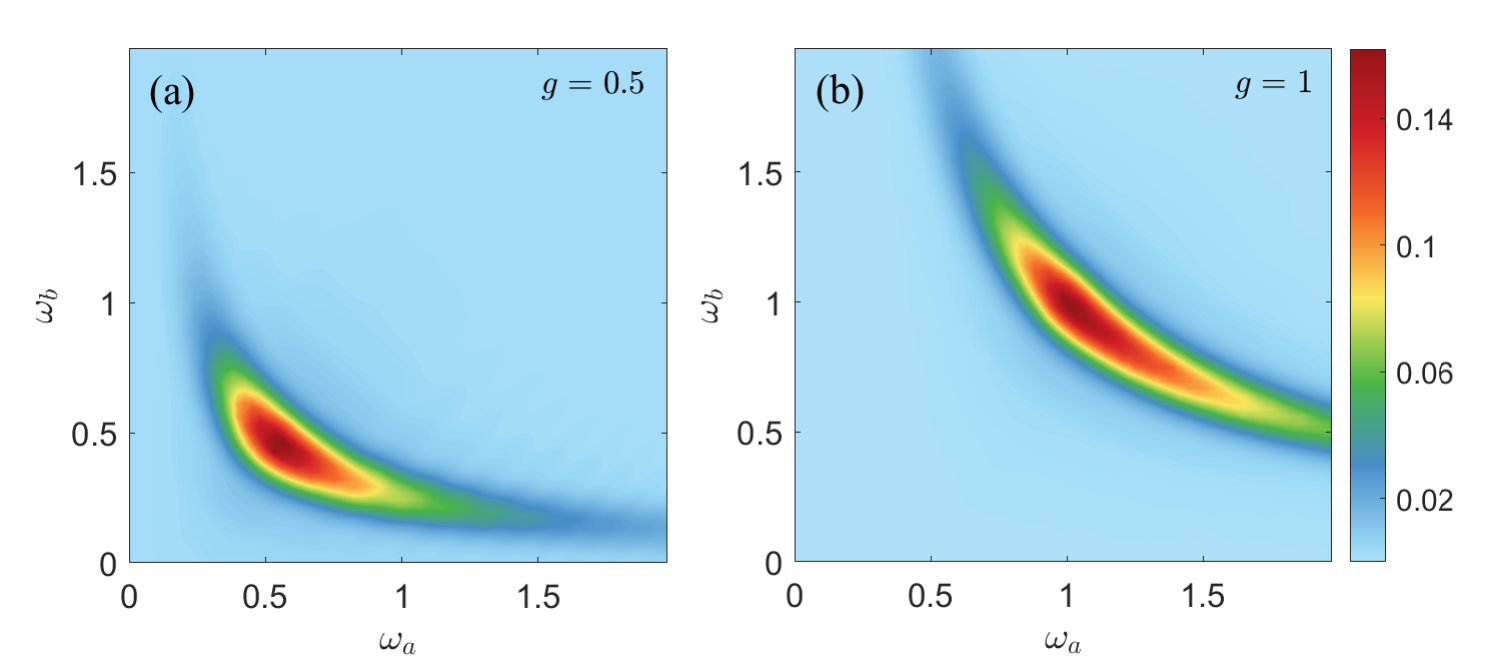}
\caption{
The impact of the detuning between $\omega_{a}$ and $\omega_{b}$ on the entanglement generation. The colors indicate the maximum entanglement generated with the given parameters. Subplots (a) and (b) represent two coupling constants  $g=0.5$ and $g=1$, respectively. The other parameters are $E=0.1$,$\gamma_s=0.005$. }
\label{fig3}
\end{figure}

Besides the impact of coupling strength $g$, we are also interested in the detuning of the Transmon from the driving field $\omega_{b}$ and the detuning of the cavity from the driving field $\omega_{a}$. Here, we compare and observe two cases with coupling constants $g=0.5$ and $g=1$, respectively. Figure \ref{fig3} shows how does the maximum $E_{N}$ behave when the detunings are different. From Figure \ref{fig3}(a), it can be observed that at $g=0.5$, the maximum entanglement occurs at a detuning of 0.5, and as the detuning changes, it significantly suppresses system entanglement. In Figure \ref{fig3}(b), for $g=1$, different levels of entanglement are observed for various detunings, with maximum entanglement occurring at a detuning of 1; as the detuning changes, its inhibitory effect on system entanglement strengthens. It can be concluded that the choice of detuning $\omega_{b}$ has a significant impact on system entanglement. These phenomena indicate that when the frequency of the cavity field precisely matches the energy level difference of the Transmon qubit (i.e., when $\omega_a$ reaches a specific value), the coupling constant $g$ maximizes the energy transfer between them, thereby effectively promoting the generation and maintenance of quantum entanglement. When $\omega_a$ deviates from the ideal value, the degree of entanglement decreases. This is because the system does not reach a resonant state, leading to suboptimal interaction efficiency between the cavity field and the Transmon qubit and the inability to stably maintain a highly entangled quantum state.

\begin{figure}[t]
\centering
\includegraphics[width=1.00\linewidth]{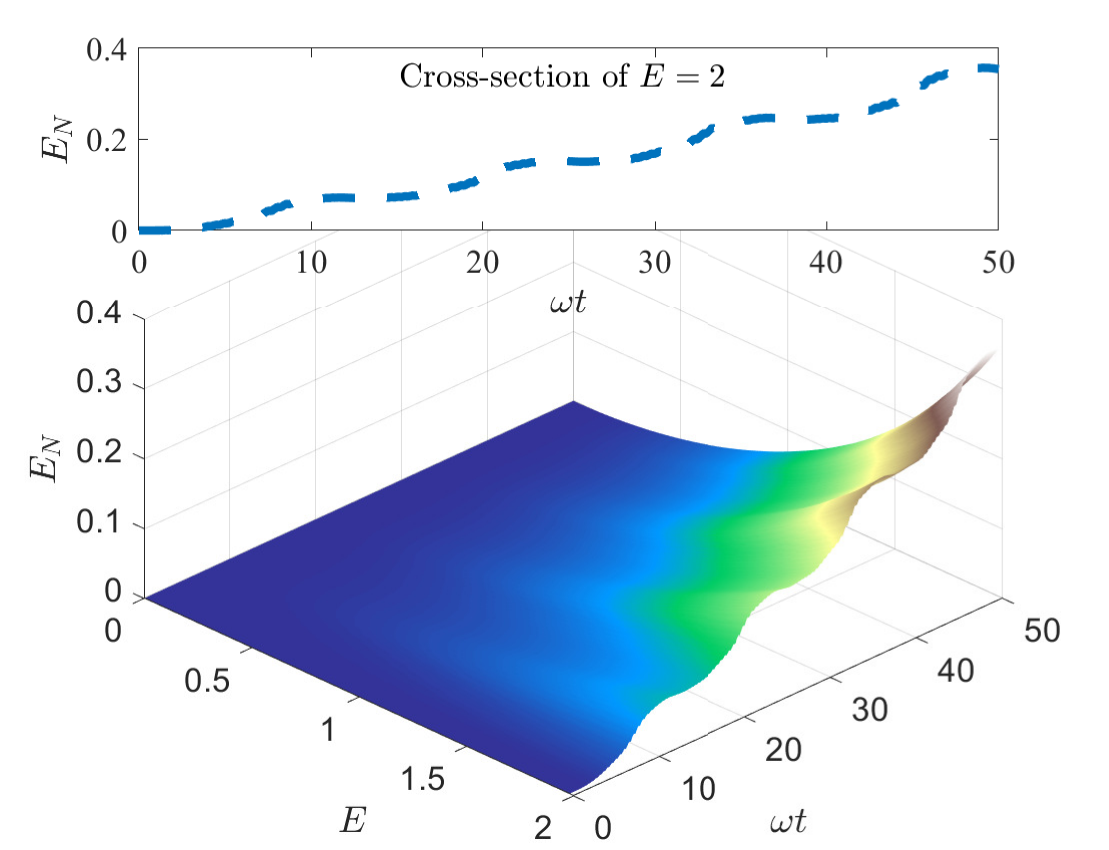}
\caption{Entanglement generation for different driving field strengths $E$. The parameters are $\omega_{b}=\omega_{a}=\omega=1$, $g=1$, and $\gamma_s=0.005$.}
\label{fig4}
\end{figure}

Last but not least, the driving field strength $E$ is also a crucial parameter to the entanglement generation. Figure \ref{fig4} shows the evolution of $E_{N}$ over time for different values of $E$. From Figure \ref{fig4}, it can be observed that as $E$ increases, the entanglement generation time shortens and the maximum value of entanglement also increases. In a separate simulation shown in the figure, when $E$ is set to 2, it can be seen that there are fluctuations in entanglement during its evolution. This phenomenon indicates that , when the driving field strength matches certain inherent characteristics of the system, such as energy level differences or resonance conditions, the entanglement can reach its maximum value. However, due to the presence of non-linear terms such as $-\frac{{E_{c}}}{2}{{\hat{b}}^{\dagger}}{{\hat{b}}^{\dagger}}\hat{b}\hat{b}$ from Transmon and the system's decoherence process may lead to a reduction in entanglement. This behavior is crucial for understanding the dynamic properties of quantum systems under external stimulation, particularly in the fields of quantum information processing and quantum communication.


\begin{figure}[t]
\centering
\includegraphics[width=1.00\linewidth]{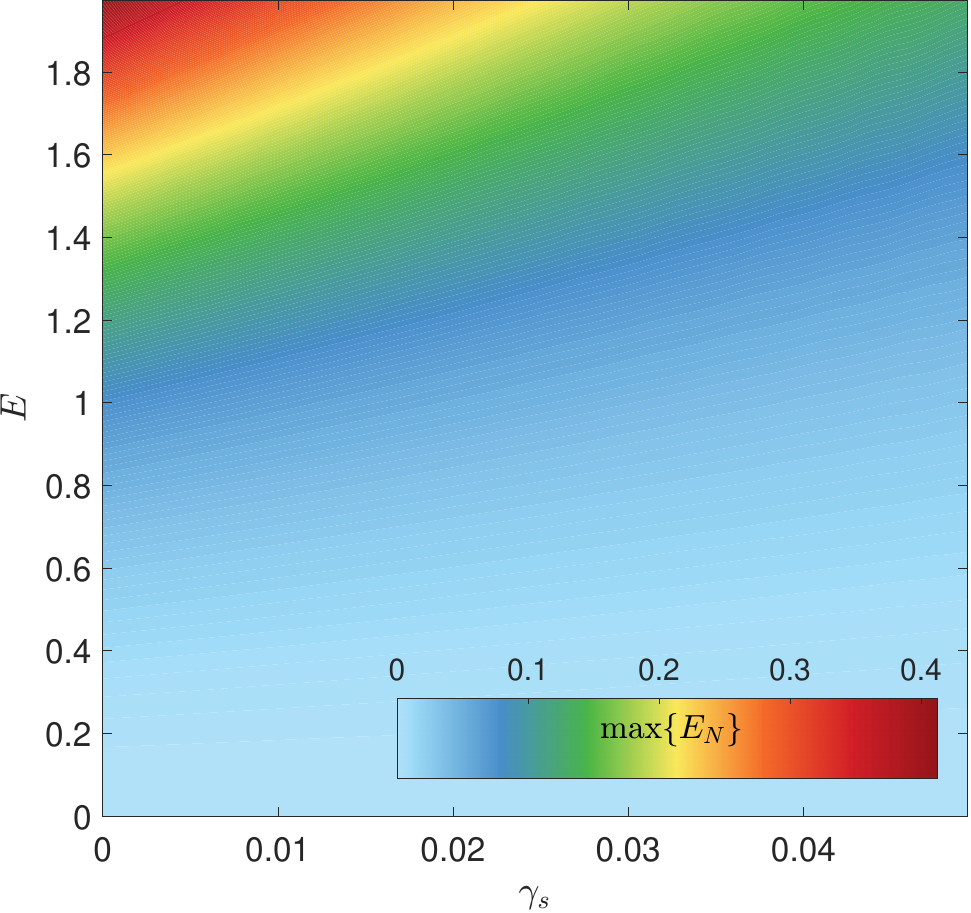}
\caption{Competition between the driving field $E$ and the dissipation rate $\gamma_s$. The time evolution of system $E_N$ is plotted for different $E$ and $\gamma_s$, with $\omega_b=1$, $\omega_a=1$, $g=0.5$.}
\label{fig5}
\end{figure}

Figure \ref{fig5} illustrates the impact of different decoherence rates $\gamma_s$ and driving field strengths $E$ on the maximum entanglement $E_N$ of the system. The numerical results show that the driving field is helpful to the entanglement while the decoherence rate is harmful to the entanglement. The maximum entanglement that can be achieved (indicated by the color in the plot) is the result of a competition between $E$ and $\gamma_s$. It is worth to note that for a specific driving field strength $E$, there is threshold for decoherence rate $\gamma_s$ to generate finite entanglement. When $\gamma_s$ exceed the threshold, the entanglement generation is zero. This phenomenon is consistent with the disruption of the system's state coherence during the quantum decoherence process. Quantum entanglement, as a fragile quantum resource, will be destroyed when the system interacts with an external environment. As quantum systems almost inevitably interact with external environments, their coherence gradually diminishes. An increase in the decoherence rate $\gamma_s$ indicates stronger coupling of the quantum system with the environment, leading to more significant perturbations of the quantum state by the environment, and consequently, a decrease in maximum entanglement. The reduction in entanglement reveals that the system's quantum state is progressively transitioning to a classical state, with its quantum characteristics gradually diminishing.

\section{Conclusion}

This article mainly investigates the entanglement between a Transmon and a cavity with an initial state of vacuum by calculating the logarithmic negativity. The effects of the driving field strength $E$, coupling constant $g$, and detuning $\omega$ on the entanglement of the system are studied. It is found that when the detunings $\omega_{a}$ and $\omega_{b}$ are equal, the system reaches maximum entanglement. The detuning has a significant inhibitory effect on system entanglement. Under resonance conditions, the entanglement between Transmon and cavity increases significantly; while under non-resonance conditions, the entanglement weakens. The coupling constant $g$ also affects system entanglement, as increasing it within a certain range can noticeably enhance entanglement. Increasing the driving field strength not only shortens the time for achieving high levels of entanglement but also increases maximum entanglement. This system exhibits high controll ability and experimental feasibility.

Transmon qubits have been widely applied in quantum computing and quantum communication due to their design flexibility and strong coupling with microwave resonators. Our model and analysis further confirm that through appropriate parameter settings and control, generating entangled states is entirely feasible and can be accurately measured in future experiments\cite{lee9,rea6,kir3}.

Looking ahead, our research has several possible directions for future development. Firstly, we can further consider the practical effects of noise and damping to obtain a more realistic model. Secondly, we can explore a wider range of parameter values to uncover other potential entanglement mechanisms. Lastly, we can attempt to develop new methods \cite{gen9,ian8,ham9,wal0,zho1,may8} for quantifying entanglement in order to provide a deeper understanding.

In general, our research provides a powerful framework for understanding and achieving entanglement between Transmon and cavity fields. The interaction model between Transmon and cavity fields presents many technical challenges in practical applications, such as realizing high-fidelity quantum gate operations, effectively maintaining the stability of Transmon, and reducing errors \cite{huan9,gen8,fer6,ved4}. However, with the continuous advancement of quantum technology, these challenges are gradually being overcome. As quantum technology continues to evolve, the potential of the interaction model between Transmon and cavity fields will be more widely realized and applied.


\appendix

\section{Expectation of collective operators with time evolution}

In Appendix A, we have deliberately listed the detailed Hamiltonian expressions of the system model and the relevant quantum Langevin equations here in order to make the main document more concise and easy to understand.

\begin{eqnarray}
\begin{aligned}
\frac{\left \langle \hat{O} \right \rangle }{dt}=Tr[\hat{O}\Dot{\rho}]
\end{aligned}
\end{eqnarray}
\begin{eqnarray}
\begin{aligned}
\Dot{\rho}&=-i[\hat{H},\rho]+\gamma_{s}\mathcal{D}[\hat{a}].
\end{aligned}
\end{eqnarray}
where $\mathcal{D}[\hat{A}]\rho=\hat{A}\rho\hat{A}^{\dagger}-\frac{1}{2}(\hat{A}^{\dagger}\hat{A}\rho+\rho\hat{A}^{\dagger}\hat{A})$ and the parameter $\gamma_{s}$  represents the loss rate of cavity photons.

\begin{flalign}
&\begin{aligned}
     & \frac{d}{dt}\left\langle \hat{a}\right\rangle =-i{{\omega}_{a}}\left\langle \hat{a}\right\rangle -ig\left\langle \hat{b}\right\rangle +E-\frac{\gamma_{s}}{2}\left\langle \hat{a}\right\rangle
       \end{aligned}\\
 &\begin{aligned}
     & \frac{d}{dt}\left\langle {{\hat{a}}^{\dagger}}\right\rangle =i{{\omega}_{a}}+ig\left\langle \hat{b}\right\rangle +E-\frac{\gamma_{s}}{2}\left\langle {{\hat{a}}^{\dagger}}\right\rangle
       \end{aligned}\\
 &\begin{aligned}
     & \frac{d}{dt}\left\langle \hat{b}\right\rangle =-i{{\omega}_{b}}\left\langle \hat{b}\right\rangle -ig\left\langle \hat{a}\right\rangle
       \end{aligned}\\
 &\begin{aligned}
         & \frac{d}{dt}\left\langle {{\hat{b}}^{\dagger}}\right\rangle =i{{\omega}_{b}}\left\langle {{\hat{b}}^{\dagger}}\right\rangle +ig\left\langle {{\hat{a}}^{\dagger}}\right\rangle
   \end{aligned}\\
 & \begin{aligned}
      \frac{d}{dt}\left\langle {{\hat{a}}^{2}}\right\rangle =&-2i{{\omega}_{a}}\left\langle {{\hat{a}}^{2}}\right\rangle -2ig\left\langle \hat{a}\hat{b}\right\rangle +2E\left\langle \hat{a}\right\rangle \\
         &-\gamma_{s}\left\langle {{\hat{a}}^{2}}\right\rangle
  \end{aligned}\\
& \begin{aligned}
          \frac{d}{dt}\left\langle {{\hat{a}}^{{{\dagger}^{2}}}}\right\rangle =&2i{{\omega}_{a}}\left\langle {{\hat{a}}^{{{\dagger}^{2}}}}\right\rangle +2ig\left\langle {{\hat{a}}^{\dagger}}{{\hat{b}}^{\dagger}}\right\rangle +2E\left\langle {{\hat{a}}^{\dagger}}\right\rangle \\
          &-\gamma_{s}\left\langle {{\hat{a}}^{{{\dagger}^{2}}}}\right\rangle\\
  \end{aligned}&&\\
 &\begin{aligned}
          \frac{d}{dt}\left\langle {{\hat{b}}^{2}}\right\rangle =-2i{{\omega}_{b}}\left\langle {{\hat{b}}^{2}}\right\rangle -2ig\left\langle \hat{a}\hat{b}\right\rangle\\
          \end{aligned}&&\\
& \begin{aligned}
          \frac{d}{dt}\left\langle {{\hat{b}}^{{{\dagger}^{2}}}}\right\rangle =&2i{{\omega}_{b}}\left\langle {{\hat{b}}^{{{\dagger}^{2}}}}\right\rangle +2ig\left\langle {{\hat{a}}^{\dagger}}{{\hat{b}}^{\dagger}}\right\rangle\\
  \end{aligned}&&\\
& \begin{aligned}
          \frac{d}{dt}\left\langle \hat{a}\hat{b}\right\rangle =&-i{{\omega}_{a}}\left\langle \hat{a}\hat{b}\right\rangle -ig\left\langle {{\hat{b}}^{2}}\right\rangle +E\left\langle \hat{b}\right\rangle \\
          &-i{{\omega}_{b}}\left\langle \hat{a}\hat{b}\right\rangle -ig\left\langle {{\hat{a}}^{2}}\right\rangle-\frac{\gamma_{s}}{2}\left\langle \hat{a}\hat{b}\right\rangle \\
  \end{aligned}&&\\
&\begin{aligned}
    \frac{d}{dt}\left\langle {{\hat{a}}^{\dagger}}{{\hat{b}}^{\dagger}}\right\rangle =& i{{\omega}_{a}}\left\langle {{\hat{a}}^{\dagger}}{{\hat{b}}^{\dagger}}\right\rangle +ig\left\langle {{\hat{b}}^{{{\dagger}^{2}}}}\right\rangle +E\left\langle {{\hat{b}}^{\dagger}}\right\rangle \\
    &+i{{\omega}_{b}}\left\langle {{\hat{a}}^{\dagger}}{{\hat{b}}^{\dagger}}\right\rangle +ig\left\langle {{\hat{a}}^{{{\dagger}^{2}}}}\right\rangle -\frac{\gamma_{s}}{2}\left\langle {{\hat{a}}^{\dagger}}{{\hat{b}}^{\dagger}}\right\rangle
\end{aligned} &&
  \end{flalign}
  \begin{flalign}
&\begin{aligned}
     \frac{d}{dt}\left\langle {{\hat{a}}^{\dagger}}\hat{b}\right\rangle =& i{{\omega}_{a}}\left\langle {{\hat{a}}^{\dagger}}\hat{b}\right\rangle +ig\left\langle {{\hat{b}}^{\dagger}}\hat{b}\right\rangle +E\left\langle \hat{b}\right\rangle \\
     &-i{{\omega}_{b}}\left\langle {{\hat{a}}^{\dagger}}\hat{b}\right\rangle -ig\left\langle {{\hat{a}}^{\dagger}}\hat{a}\right\rangle -\frac{\gamma_{s}}{2}\left\langle {{\hat{a}}^{\dagger}}\hat{b}\right\rangle
\end{aligned} && \\
&\begin{aligned}
      \frac{d}{dt}\left\langle {\hat{a}{\hat{b}}^{\dagger}}\right\rangle =& -i{{\omega}_{a}}\left\langle {\hat{a}{\hat{b}}^{\dagger}}\right\rangle -ig\left\langle {{\hat{b}}^{\dagger}\hat{b}}\right\rangle +E\left\langle {{\hat{b}}^{\dagger}}\right\rangle \\
      &+i{{\omega}_{b}}\left\langle {\hat{a}{\hat{b}}^{\dagger}}\right\rangle +ig\left\langle {{\hat{a}}^{\dagger}\hat{a}}\right\rangle -\frac{\gamma_{s}}{2}\left\langle {\hat{a}{\hat{b}}^{\dagger}}\right\rangle
 \end{aligned} &&\\
 & \begin{aligned}
           \frac{d}{dt}\left\langle {{\hat{a}}^{\dagger}}\hat{a}\right\rangle =&ig\left\langle \hat{a}{{\hat{b}}^{\dagger}}\right\rangle -ig\left\langle {{\hat{a}}^{\dagger}}\hat{b}\right\rangle  \\
          &+E\left\langle \hat{a}\right\rangle +E\left\langle {{\hat{a}}^{\dagger}}\right\rangle-\gamma_{s}\left\langle {{\hat{a}}^{\dagger}}\hat{a}\right\rangle\\
   \end{aligned}&&\\
& \begin{aligned}
          \frac{d}{dt}\left\langle \hat{a}{{\hat{a}}^{\dagger}}\right\rangle =&ig\left\langle \hat{a}{{\hat{b}}^{\dagger}}\right\rangle -ig\left\langle {{\hat{a}}^{\dagger}}\hat{b}\right\rangle \\
          &+E\left\langle \hat{a}\right\rangle +E\left\langle {{\hat{a}}^{\dagger}}\right\rangle-\gamma_{s}\left\langle {{\hat{a}}^{\dagger}}\hat{a}\right\rangle\\
  \end{aligned}&&\\
& \begin{aligned}
          \frac{d}{dt}\left\langle {{\hat{b}}^{\dagger}}\hat{b}\right\rangle =-ig\left\langle \hat{a}{{\hat{b}}^{\dagger}}\right\rangle +ig\left\langle {{\hat{a}}^{\dagger}}\hat{b}\right\rangle
 \end{aligned}&&\\
 &\begin{aligned}
          \frac{d}{dt}\left\langle \hat{b}{{\hat{b}}^{\dagger}}\right\rangle =-ig\left\langle \hat{a}{{\hat{b}}^{\dagger}}\right\rangle +ig\left\langle {{\hat{a}}^{\dagger}}\hat{b}\right\rangle
\end{aligned}&&
\end{flalign}

We have neglected the impact brought by higher-order terms.We can rewrite the above formula in matrix form, which is concise and clear, and helps us understand the dynamic characteristics of the system.

\begin{eqnarray}
\dot{f}\left(t\right)=Af\left(t\right)+\chi\left(t\right).
\end{eqnarray}

The elements in matrix A are
$A_{2,4}=A_{4,2}=A_{10,6}=A_{10,8}=A_{11,15}=A_{12,13}=A_{13,12}=A_{14,12}=A_{15,11}=A_{16,11}=ig$,
$A_{1,3}=A_{3,1}=A_{9,5}=A_{9,7}=A_{11,13}=A_{12,15}=A_{13,11}=A_{14,11}=A_{15,12}=A_{16,12}=-ig$,
$A_{6,10}=A_{8,10}=2ig$,
$A_{5,9}=A_{7,9}=-2ig$,
$A_{9,3}=A_{10,4}=A_{11,3}=A_{12,4}=A_{13,1}=A_{13,2}=A_{14,1}=A_{14,2}=E$,$A_{5,1}=A_{6,2}=2E$,
$A_{1,1}=-i\omega_a-\frac{\gamma_s}{2}$,
$A_{2,2}=i\omega_a-\frac{\gamma_s}{2}$,
$A_{5,5}=-2i\omega_a-\gamma_s$,
$A_{6,6}=2i\omega_a-\gamma_s$,
$A_{9,9}=-i(\omega_a+\omega_b)-\frac{\gamma_s}{2}$,
$A_{10,10}=i(\omega_a+\omega_b)-\frac{\gamma_s}{2}$,
$A_{11,11}=i(\omega_a-\omega_b)-\frac{\gamma_s}{2}$,
$A_{12,12}=-i(\omega_a-\omega_b)-\frac{\gamma_s}{2}$
$A_{13,13}=A_{14,13}=-\gamma_s$.All other elements in matrix A are 0.


This way of expression is concise and clear, which helps us understand
the dynamic characteristics of the system.
{{
\section{Spectrum and dynamics of beams at the interface of two arrays in the trivial phase}

The covariance matrix $\Gamma$ is essentially a $4\times4$ matrix, which we have divided into $2\times2$ sub-matrices. The element in the $i^{th}$ row and $j^{th}$ column is represented as:

\begin{eqnarray}
{{\Gamma}_{ij}}=\frac{1}{2}\left\langle {{x}_{i}}{{x}_{j}}+{{x}_{j}}{{x}_{i}}\right\rangle -\left\langle {{x}_{i}}\right\rangle \left\langle {{x}_{j}}\right\rangle .
\end{eqnarray}

\noindent where $x_{1}=\frac{\hat{a}+\hat{a}^{\dagger}}{\sqrt{2}}$, $x_{2}=\frac{\hat{a}-\hat{a}^{\dagger}}{\sqrt{2}i}$, $x_{3}=\frac{\hat{b}+\hat{b}^{\dagger}}{\sqrt{2}}$, $x_{4}=\frac{\hat{b}-\hat{b}^{\dagger}}{\sqrt{2}i}$.
}}
\begin{flalign}
&\begin{aligned}
      {{\Gamma}_{11}}=&\frac{1}{2}\left(\left\langle \hat{a}\hat{a}\right\rangle +\left\langle \hat{a}{{\hat{a}}^{\dagger}}\right\rangle +\left\langle {{\hat{a}}^{\dagger}}\hat{a}\right\rangle +\left\langle {{\hat{a}}^{\dagger}}{{\hat{a}}^{\dagger}}\right\rangle \right)\\
      &-\frac{1}{2}\left({{\left\langle {\hat{a}}\right\rangle }^{2}}+\left\langle {\hat{a}}\right\rangle \left\langle {{\hat{a}}^{\dagger}}\right\rangle +\left\langle {{\hat{a}}^{\dagger}}\right\rangle \left\langle {\hat{a}}\right\rangle +{{\left\langle {{\hat{a}}^{\dagger}}\right\rangle }^{2}}\right)\\
\end{aligned}\\
&\begin{aligned}
     {{\Gamma}_{12}}=&\frac{1}{2i}\left(\left\langle \hat{a}\hat{a}\right\rangle -\left\langle {{\hat{a}}^{\dagger}}{{\hat{a}}^{\dagger}}\right\rangle \right)\\
     &-\frac{1}{2i}\left({{\left\langle {\hat{a}}\right\rangle }^{2}}-\left\langle {\hat{a}}\right\rangle \left\langle {{\hat{a}}^{\dagger}}\right\rangle +\left\langle {{\hat{a}}^{\dagger}}\right\rangle \left\langle {\hat{a}}\right\rangle -{{\left\langle {{\hat{a}}^{\dagger}}\right\rangle }^{2}}\right)\\
 \end{aligned}\\
&\begin{aligned}
     {{\Gamma}_{13}}=& \frac{1}{2}\left(\left\langle \hat{a}\hat{b}\right\rangle +\left\langle \hat{a}{{\hat{b}}^{\dagger}}\right\rangle +\left\langle {{\hat{a}}^{\dagger}}\hat{b}\right\rangle +\left\langle {{\hat{a}}^{\dagger}}{{\hat{b}}^{\dagger}}\right\rangle \right)\\
     &-\frac{1}{2}\left(\left\langle {\hat{a}}\right\rangle \left\langle {\hat{b}}\right\rangle +\left\langle {\hat{a}}\right\rangle \left\langle {{\hat{b}}^{\dagger}}\right\rangle +\left\langle {{\hat{a}}^{\dagger}}\right\rangle \left\langle {\hat{b}}\right\rangle +\left\langle {{\hat{a}}^{\dagger}}\right\rangle \left\langle {{\hat{b}}^{\dagger}}\right\rangle \right)\\
 \end{aligned}\\
&\begin{aligned}
       {{\Gamma}_{14}}=&\frac{1}{2i}\left(\left\langle \hat{a}\hat{b}\right\rangle +\left\langle {{\hat{a}}^{\dagger}}\hat{b}\right\rangle -\left\langle \hat{a}{{\hat{b}}^{\dagger}}\right\rangle -\left\langle {{\hat{a}}^{\dagger}}{{\hat{b}}^{\dagger}}\right\rangle \right)\\
       &-\frac{1}{2i}\left(\left\langle {\hat{a}}\right\rangle \left\langle {\hat{b}}\right\rangle -\left\langle {\hat{a}}\right\rangle \left\langle {{\hat{b}}^{\dagger}}\right\rangle +\left\langle {{\hat{a}}^{\dagger}}\right\rangle \left\langle {\hat{b}}\right\rangle -\left\langle {{\hat{a}}^{\dagger}}\right\rangle \left\langle {{\hat{b}}^{\dagger}}\right\rangle \right)\\
 \end{aligned}\\
&\begin{aligned}
     {{\Gamma}_{21}}=& \frac{1}{2i}\left(\left\langle \hat{a}\hat{a}\right\rangle -\left\langle {{\hat{a}}^{\dagger}}{{\hat{a}}^{\dagger}}\right\rangle \right)\\
     &-\frac{1}{2i}\left({{\left\langle {\hat{a}}\right\rangle }^{2}}+\left\langle {\hat{a}}\right\rangle \left\langle {{\hat{a}}^{\dagger}}\right\rangle -\left\langle {{\hat{a}}^{\dagger}}\right\rangle \left\langle {\hat{a}}\right\rangle -{{\left\langle {{\hat{a}}^{\dagger}}\right\rangle }^{2}}\right)\\
 \end{aligned}\\
    \end{flalign}
 \begin{flalign}
&\begin{aligned}
     {{\Gamma}_{22}}=& \frac{1}{2}\left(\left\langle \hat{a}\hat{a}\right\rangle -\left\langle \hat{a}{{\hat{a}}^{\dagger}}\right\rangle -\left\langle {{\hat{a}}^{\dagger}}\hat{a}\right\rangle +\left\langle {{\hat{a}}^{\dagger}}{{\hat{a}}^{\dagger}}\right\rangle \right)\\
     &+\frac{1}{2}\left({{\left\langle {\hat{a}}\right\rangle }^{2}}-\left\langle {\hat{a}}\right\rangle \left\langle {{\hat{a}}^{\dagger}}\right\rangle -\left\langle {{\hat{a}}^{\dagger}}\right\rangle \left\langle {\hat{a}}\right\rangle +{{\left\langle {{\hat{a}}^{\dagger}}\right\rangle }^{2}}\right)\\
 \end{aligned}\\
 &\begin{aligned}
    {{\Gamma}_{23}}=& \frac{1}{2i}\left(\left\langle \hat{a}\hat{b}\right\rangle -\left\langle {{\hat{a}}^{\dagger}}\hat{b}\right\rangle +\left\langle \hat{a}{{\hat{b}}^{\dagger}}\right\rangle -\left\langle {{\hat{a}}^{\dagger}}{{\hat{b}}^{\dagger}}\right\rangle \right)\\
    &-\frac{1}{2i}\left(\left\langle {\hat{a}}\right\rangle \left\langle {\hat{b}}\right\rangle +\left\langle {\hat{a}}\right\rangle \left\langle {{\hat{b}}^{\dagger}}\right\rangle -\left\langle {{\hat{a}}^{\dagger}}\right\rangle \left\langle {\hat{b}}\right\rangle -\left\langle {{\hat{a}}^{\dagger}}\right\rangle \left\langle {{\hat{b}}^{\dagger}}\right\rangle \right)\\
 \end{aligned}\\
&\begin{aligned}
      {{\Gamma}_{24}}=& \frac{1}{2}\left(\left\langle \hat{a}\hat{b}\right\rangle -\left\langle {{\hat{a}}^{\dagger}}\hat{b}\right\rangle -\left\langle \hat{a}{{\hat{b}}^{\dagger}}\right\rangle +\left\langle {{\hat{a}}^{\dagger}}{{\hat{b}}^{\dagger}}\right\rangle \right)\\
      &+\frac{1}{2}\left(\left\langle {\hat{a}}\right\rangle \left\langle {\hat{b}}\right\rangle -\left\langle {\hat{a}}\right\rangle \left\langle {{\hat{b}}^{\dagger}}\right\rangle -\left\langle {{\hat{a}}^{\dagger}}\right\rangle \left\langle {\hat{b}}\right\rangle +\left\langle {{\hat{a}}^{\dagger}}\right\rangle \left\langle {{\hat{b}}^{\dagger}}\right\rangle \right)\\
 \end{aligned}\\
&\begin{aligned}
    {{\Gamma}_{31}}=& \frac{1}{2}\left(\left\langle \hat{a}\hat{b}\right\rangle +\left\langle \hat{a}{{\hat{b}}^{\dagger}}\right\rangle +\left\langle {{\hat{a}}^{\dagger}}\hat{b}\right\rangle +\left\langle {{\hat{a}}^{\dagger}}{{\hat{b}}^{\dagger}}\right\rangle \right)\\
    &-\frac{1}{2}\left(\left\langle {\hat{b}}\right\rangle \left\langle {\hat{a}}\right\rangle +\left\langle {\hat{b}}\right\rangle \left\langle {{\hat{a}}^{\dagger}}\right\rangle +\left\langle {{\hat{b}}^{\dagger}}\right\rangle \left\langle {\hat{a}}\right\rangle +\left\langle {{\hat{b}}^{\dagger}}\right\rangle \left\langle {{\hat{a}}^{\dagger}}\right\rangle \right)\\
 \end{aligned}\\
&\begin{aligned}
     {{\Gamma}_{32}}=& \frac{1}{2i}\left(\left\langle \hat{a}\hat{b}\right\rangle -\left\langle {{\hat{a}}^{\dagger}}\hat{b}\right\rangle +\left\langle \hat{a}{{\hat{b}}^{\dagger}}\right\rangle -\left\langle {{\hat{a}}^{\dagger}}{{\hat{b}}^{\dagger}}\right\rangle \right)\\
     &-\frac{1}{2i}\left(\left\langle {\hat{b}}\right\rangle \left\langle {\hat{a}}\right\rangle -\left\langle {\hat{b}}\right\rangle \left\langle {{\hat{a}}^{\dagger}}\right\rangle +\left\langle {{\hat{b}}^{\dagger}}\right\rangle \left\langle {\hat{a}}\right\rangle -\left\langle {{\hat{b}}^{\dagger}}\right\rangle \left\langle {{\hat{a}}^{\dagger}}\right\rangle \right)\\
 \end{aligned}\\
&\begin{aligned}
     {{\Gamma}_{33}}=& \frac{1}{2}\left(\left\langle \hat{b}\hat{b}\right\rangle +\left\langle \hat{b}{{\hat{b}}^{\dagger}}\right\rangle +\left\langle {{\hat{b}}^{\dagger}}\hat{b}\right\rangle +\left\langle {{\hat{b}}^{\dagger}}{{\hat{b}}^{\dagger}}\right\rangle \right)\\
     &-\frac{1}{2}\left({{\left\langle {\hat{b}}\right\rangle }^{2}}+\left\langle {\hat{b}}\right\rangle \left\langle {{\hat{b}}^{\dagger}}\right\rangle +\left\langle {{\hat{b}}^{\dagger}}\right\rangle \left\langle {\hat{b}}\right\rangle +{{\left\langle {{\hat{b}}^{\dagger}}\right\rangle }^{2}}\right)\\
 \end{aligned}
\end{flalign}
\begin{flalign}
&\begin{aligned}
      {{\Gamma}_{34}}=& \frac{1}{2i}\left(\left\langle \hat{b}\hat{b}\right\rangle -\left\langle {{\hat{b}}^{\dagger}}{{\hat{b}}^{\dagger}}\right\rangle \right)\\
      &-\frac{1}{2i}\left(\left\langle {\hat{b}}\right\rangle \left\langle {\hat{b}}\right\rangle -\left\langle {\hat{b}}\right\rangle \left\langle {{\hat{b}}^{\dagger}}\right\rangle +\left\langle {{\hat{b}}^{\dagger}}\right\rangle \left\langle {\hat{b}}\right\rangle -\left\langle {{\hat{b}}^{\dagger}}\right\rangle \left\langle {{\hat{b}}^{\dagger}}\right\rangle \right)\\
 \end{aligned}\\
&\begin{aligned}
    {{\Gamma}_{41}}=& \frac{1}{2i}\left(\left\langle \hat{a}\hat{b}\right\rangle +\left\langle {{\hat{a}}^{\dagger}}\hat{b}\right\rangle -\left\langle \hat{a}{{\hat{b}}^{\dagger}}\right\rangle -\left\langle {{\hat{a}}^{\dagger}}{{\hat{b}}^{\dagger}}\right\rangle \right)\\
    &-\frac{1}{2i}\left(\left\langle {\hat{b}}\right\rangle \left\langle {\hat{a}}\right\rangle +\left\langle {\hat{b}}\right\rangle \left\langle {{\hat{a}}^{\dagger}}\right\rangle -\left\langle {{\hat{b}}^{\dagger}}\right\rangle \left\langle {\hat{a}}\right\rangle -\left\langle {{\hat{b}}^{\dagger}}\right\rangle \left\langle {{\hat{a}}^{\dagger}}\right\rangle \right)\\
 \end{aligned}\\
&\begin{aligned}
      {{\Gamma}_{42}}=& -\frac{1}{2}\left(\left\langle \hat{a}\hat{b}\right\rangle -\left\langle {{\hat{a}}^{\dagger}}\hat{b}\right\rangle -\left\langle \hat{a}{{\hat{b}}^{\dagger}}\right\rangle +\left\langle {{\hat{a}}^{\dagger}}{{\hat{b}}^{\dagger}}\right\rangle \right)\\
      &+\frac{1}{2}\left(\left\langle {\hat{b}}\right\rangle \left\langle {\hat{a}}\right\rangle -\left\langle {\hat{b}}\right\rangle \left\langle {{\hat{a}}^{\dagger}}\right\rangle -\left\langle {{\hat{b}}^{\dagger}}\right\rangle \left\langle {\hat{a}}\right\rangle +\left\langle {{\hat{b}}^{\dagger}}\right\rangle \left\langle {{\hat{a}}^{\dagger}}\right\rangle \right)\\
 \end{aligned}
   \end{flalign}

  \begin{flalign}
&\begin{aligned}
    {{\Gamma}_{43}}=& \frac{1}{2i}\left(\left\langle \hat{b}\hat{b}\right\rangle -\left\langle {{\hat{b}}^{\dagger}}{{\hat{b}}^{\dagger}}\right\rangle \right)\\
    &-\frac{1}{2i}\left({{\left\langle {\hat{b}}\right\rangle }^{2}}+\left\langle {\hat{b}}\right\rangle \left\langle {{\hat{b}}^{\dagger}}\right\rangle -\left\langle {{\hat{b}}^{\dagger}}\right\rangle \left\langle {\hat{b}}\right\rangle -{{\left\langle {{\hat{b}}^{\dagger}}\right\rangle }^{2}}\right)\\
 \end{aligned}\\
&\begin{aligned}
      {{\Gamma}_{44}}=& -\frac{1}{2}\left(\left\langle \hat{b}\hat{b}\right\rangle -\left\langle \hat{b}{{\hat{b}}^{\dagger}}\right\rangle -\left\langle {{\hat{b}}^{\dagger}}\hat{b}\right\rangle +\left\langle {{\hat{b}}^{\dagger}}{{\hat{b}}^{\dagger}}\right\rangle \right)\\
      &+\frac{1}{2}\left({{\left\langle {\hat{b}}\right\rangle }^{2}}-\left\langle {\hat{b}}\right\rangle \left\langle {{\hat{b}}^{\dagger}}\right\rangle -\left\langle {{\hat{b}}^{\dagger}}\right\rangle \left\langle {\hat{b}}\right\rangle +{{\left\langle {{\hat{b}}^{\dagger}}\right\rangle }^{2}}\right)\\
 \end{aligned}
\end{flalign}




\end{document}